\documentclass[prx,aps,reprint,showpacs,amsfonts,amsmath,floatfix
,superscriptaddress,longbibliography]{revtex4-2}

\usepackage[T1]{fontenc}
\usepackage[svgnames]{xcolor}

\usepackage[colorlinks=true,        
            allcolors = black,  
            citecolor=DarkBlue, 
            linkcolor=DarkBlue,
            urlcolor=DarkBlue
        ]{hyperref}

\usepackage{amsmath, amsfonts, amssymb, amsthm, bbm}
\usepackage{dsfont}
\usepackage{bm} 
\usepackage{mathtools}
\usepackage{physics}
\usepackage{bm}
\usepackage{float}
\usepackage{dcolumn}
\usepackage{soul}

\usepackage{tikz}
\usetikzlibrary{quantikz}
\usepackage{graphicx}
\usepackage{dcolumn}
\usepackage[acronym]{glossaries}
\usepackage{array}   
\usepackage{txfonts}

\newcolumntype{L}{>{$}l<{$}} 

\newacronym{vqa}{VQA}{variational quantum algorithm}
\newacronym{qaoa}{QAOA}{Quantum Approximate Optimization Algorithm}
\newacronym{vqe}{VQE}{Variational Quantum Eigensolver}
\newacronym{hva}{HVA}{Hamiltonian Variational Ansatz}
\newacronym{hea}{HEA}{Hardware-Efficient Ansatz}
\newacronym{qaa}{QAA}{Quantum Adiabatic Algorithm}
\newacronym{nisq}{NISQ}{Noisy Intermediate Scale Quantum}
\newacronym{qubo}{QUBO}{Quadratic Unconstrained Binary Optimization}

\newcommand  {\sv}  	{\bm{s}}
\newcommand	{\NH}	{N_{\text{H}}}
\newcommand	{\NHH}	{N_{\text{HH}}}
\newcommand	{\EHP}	{E_{\text{HP}}}
\newcommand{\bigO}{\mathcal{O}}
\newcommand{\EfficientSU}{hardware efficient SU(2) 2-local }

\usepackage{comment}

\begin{document}

\preprint{APS/123-QED}

\title{Designing lattice proteins with variational quantum algorithms}

\author{Hanna Linn}
\email{hannlinn@chalmers.se}
\affiliation{%
Department of Microtechnology and Nanoscience (MC2), Chalmers University of Technology, SE-412 96 G\"{o}teborg, Sweden
}%
\author{Lucas Knuthson}%
\affiliation{Computational Science for Health and Environment (COSHE), Lund University, SE-223 62 Lund, Sweden}
\author{Anders Irbäck}
\affiliation{Computational Science for Health and Environment (COSHE), Lund University, SE-223 62 Lund, Sweden}
\author{Sandipan Mohanty}
\affiliation{J\"ulich
Supercomputing Centre, Forschungszentrum J\"ulich, D-52425 J\"ulich, Germany}

\author{Laura García-Álvarez}
\author{G\"{o}ran Johansson}
\affiliation{%
Department of Microtechnology and Nanoscience (MC2), Chalmers University of Technology, SE-412 96 G\"{o}teborg, Sweden
}%

\begin{abstract}
Quantum heuristics have shown promise in solving various optimization problems, including lattice protein folding. 
Equally relevant is the inverse problem, protein design, where one seeks sequences that fold to a given target structure. The latter problem is often split into two steps: (i) searching for sequences that minimize the energy in the target structure, and (ii) testing whether the generated sequences fold to the desired structure. Here, we investigate the utility of variational quantum algorithms for the first of these two steps on today's noisy intermediate-scale quantum devices. We focus on the sequence optimization task, which is less resource-demanding than folding computations. We test the quantum approximate optimization algorithm and variants of it, with problem-informed quantum circuits, as well as the hardware-efficient ansatz, with problem-agnostic quantum circuits. While the former approach, with a careful choice of mixer, yields good results in noiseless simulations (success probability $\geq$0.95 in all instances), its performance drops drastically under noise. With the problem-agnostic circuits, which are more compatible with hardware constraints, improved performance is observed in noisy simulations, compared to their problem-informed counterparts. When running the problem-agnostic circuits on a real quantum device, with parameters taken from simulations, we obtain a significant success probability ($\geq$0.2) in most instances.

\end{abstract}

\maketitle


\section{\label{sec:intro} Introduction}

The field of quantum computing is undergoing rapid change, with significant advances made in recent years~\cite{Wang:25, Krinner:22, Acharya:23, Acharya:24}. Intense efforts are being directed toward building both analog quantum annealers and gate-based quantum hardware, with gate-based systems as the dominant model, because of its flexibility and universality.
As gate-based quantum hardware continues to evolve, \glspl{vqa} are gaining traction in fields ranging from materials science to machine learning~\cite{Cerezo:21}. Algorithms such as the \gls{qaoa} and its variants~\cite{Farhi:14, Hadfield:19}, the Hamiltonian Variational Ansatz~\cite{Wecker:15}, and the \gls{hea}~\cite{Kandala:17} have emerged as strong candidates for solving discrete optimization problems on \gls{nisq} devices~\cite{Preskill:18}. Despite the potential of quantum computing, real quantum devices are currently limited by noise and imperfections. Therefore, to make the most of the available quantum resources, these algorithms leverage parameterized quantum circuits, with parameters optimized iteratively by classical computations, to find the solution. This hybrid quantum-classical approach is key for the feasibility of these algorithms on \gls{nisq} devices; however, their practical success depends heavily on effective initialization heuristics, robust parameter optimization, and strategies for mitigating noise-induced errors~\cite{McClean:18}.

Among these challenges, parameter optimization stands out as a critical bottleneck. Variational circuits often suffer from barren plateaus and poor convergence in high-dimensional parameter landscapes~\cite{McClean:18}. To address this hurdle, transfer learning techniques---such as parameter donation between related problem instances or distributed initialization strategies---have been proposed to accelerate convergence and enhance robustness. In particular, parameter donation has shown promise in \gls{qaoa}~\cite{Langfitt:23, Montanez-Barrera:24,Galda:23,Lyngfelt:25} where reusing optimized parameters from smaller or similar instances can guide the optimization process more effectively. These strategies are especially relevant in the context of current quantum hardware, where limited coherence times and gate fidelities impose strict constraints on circuit depth and runtime~\cite{Preskill:18, Kosen:22}.

A biophysically relevant optimization problem that has been explored using digital \glspl{vqa}~\cite{fingerhuth_quantum_2018,Robert:21,Boulebnane:23,Linn:25} as well as analog quantum annealing~\cite{Perdomo-Ortiz:12,Outeiral:21,Irback:22, Irback:25} is the folding of lattice proteins, where the task is to find the minimum energy structure(s) for a given amino acid sequence. In Ref.~\cite{Irback:22}, the authors found that a field-like representation in conjunction with the ready availability of a large number of qubits in D-Wave quantum annealers allows hybrid quantum-classical sampling to compete favorably with established classical methods. However, the number of qubits and gates required to implement digital quantum approaches to this problem makes their implementation on current NISQ devices impractical beyond proof-of-concept problem sizes~\cite{Linn:24}. Another important biophysical challenge is the inverse problem~\cite{Kuhlman:03,Bhardway:16,Yang:19}, known as protein design, where one looks for sequences that fold into a given target structure. Recent years have seen great advances in protein design methods, in part based on machine learning techniques~\cite{Kuhlman:19,Cao:20}. However, computational analysis of the biophysics of protein design remains a challenge. The possibility of using quantum optimization to speed up such computations has recently been addressed, focusing on quantum annealing~\cite{Mulligan:20,Irback:24,Panizza:24} and Grover's algorithm~\cite{Khatami:23}.

In this paper, we explore the utility of \glspl{vqa} for the design of lattice proteins, through classical simulations of quantum circuits and, in selected cases, tests on quantum hardware. We are not aware of any previous work that explores the use of such quantum circuits for the protein design problem. Specifically, we consider the problem of determining amino acid sequences that minimize the energy in a given target structure, using the 2D hydrophobic/polar (HP) model of Lau and Dill~\cite{Lau:89,Yue:95} as a test bed. In this problem, the degrees of freedom are types rather than positions of the amino acids, which makes this task less resource-demanding than the folding problem. However, whether or not the generated sequences actually fold to the desired structure generally needs to be checked. The problem instances examined in this article were chosen so that this verification step is not required. Using exact results available for HP chains with lengths $N\le 30$~\cite{Irback:02,Holzgrafe:11}, we choose instances such that the sequence optimization problem has a unique solution that is also known to fold to the target structure. Our choice of simplified yet non-trivial problems with a priori known exact solutions helps us evaluate the effectiveness of nascent computational techniques such as \glspl{vqa} and identify the inherent challenges impeding their wider applicability.  

To this end, we test and compare two types of \glspl{vqa}: problem-informed \gls{qaoa} variants, which incorporate the problem structure into the quantum circuit through parametrized gates derived from the objective function; and the problem-agnostic \gls{hea} approach, for which the quantum circuit structure is problem-independent and tailored to the hardware capabilities. That is, the quantum circuit is formulated using the gate set and connectivity of the device at hand, to minimize circuit depth.
While the best of the \gls{qaoa} variants performs well in noise-free simulations (probability $\geq$0.95 to find the ground state), all variants tested prove impractical when adding noise due to the substantial circuit depths demanded. In contrast, when using \gls{hea}, our problems could also be solved with noise, especially when donating parameters between similar problem instances. When running on the IBM Torino quantum device, using pre-determined parameters taken from the simulations, two-layer \gls{hea} had a success probability $\geq$0.2 for most instances studied, with $N$ up to 29.

The structure of this paper is as follows. Section~\ref{sec:methods} reviews the formulation of the sequence optimization problem for quantum computing devices and details our methodology, including quantum circuit design and classical optimization strategies. Section~\ref{sec:results} presents our numerical and experimental results, which are further discussed in Sec.~\ref{sec:Discussion}. We conclude with a discussion and outlook in Sec.~\ref{sec:conclusion}.

\section{Methods \label{sec:methods}}

\subsection{Protein design theory}
\label{sec:protein_design_theory}

In protein design, an amino acid sequence $\sv=(s_1,\ldots,s_N)$ is sought that folds into a given target structure $C_t$.
To this end, the goal is to maximize the probability of finding the chain in the state $C_t$, given by
\begin{equation}
P_\beta(\sv)=e^{-\beta E(C_t,\sv)} \Big/ \sum_C e^{-\beta E(C,\sv)},
\label{eq:P}
\end{equation}
where $E(C_t,\sv)$ is the energy of the sequence $\sv$ in conformation $C_t$,
$\beta$ is the inverse temperature, and the sum runs over all possible structures $C$.
Although methods for this task have been developed~\cite{Irback:99,Aina:17}, maximizing $P_\beta(\sv)$ involves a generally time-consuming search in both sequence and structure spaces. Therefore, a common approach is to ignore the $\sv$ dependence of the sum in Eq.~(\ref{eq:P}) and thus minimize $E(C_t,\sv)$~\cite{Mulligan:20,Irback:24}, or introduce an approximation $e^{-\beta\tilde F(\sv)}$ of the sum and minimize $E(C_t,\sv)-\tilde F(\sv)$~\cite{Panizza:24}. In either case, an additional filtering step is needed to reject candidate sequences that have a higher probability for some other structure $C\ne C_t$.

In this work, we focus on the problem of minimizing $E(C_t,\sv)$ over $s$, for brevity referred to as sequence optimization. In all the instances studied below, the sequences generated in this way fold to the desired structure $C_t$~\cite{Irback:02,Holzgrafe:11}. 

\subsection{HP lattice proteins\label{sec:methods_HP}}

We consider the minimal two-dimensional lattice-based HP model of proteins~\cite{Lau:89}, in which the protein is represented by a self-avoiding chain of $N$ hydrophobic (H) or polar (P) beads that interact through a pairwise contact potential. A contact between two beads is said to occur
if they are nearest neighbors on the lattice but not along the chain. The energy function is defined
as $\EHP=-\NHH$, with $\NHH$ the number of HH contacts~\cite{Lau:89}. This definition
renders the formation of a hydrophobic core energetically favorable. 

The ground state, i.e., the state of minimum energy, may be degenerate or unique. On a two-dimensional square lattice, it is known from exhaustive enumerations that
about 2\% of all HP sequences with length $N \leq 30$ have a unique ground state~\cite{Irback:02, Holzgrafe:11}.
A sequence whose ground state is unique is said to design that structure. The designability of a structure is the number of sequences that design it. High designability implies mutation-tolerance and is a characteristic of real proteins. In our computations, we focus on one target structure for each chain length $N$, chosen as the most designable structure for that $N$.

Despite their simplicity, coarse-grained HP models are still relevant for qualitative insights into computationally challenging problems like liquid-liquid phase separation of intrinsically disordered proteins~\cite{Nilsson:20,Statt:20} and protein evolution modeling~\cite{Bornberg-Bauer:99,Aguirre:18}.

\subsection{HP sequence optimization in QUBO form\label{sec:methods_qubo}}

Given a target structure $C_t$, we want to find sequences $\sv$ that minimize the energy $\EHP(C_t,\sv)$, using variational quantum circuits. 
To this end, we recast the problem in \gls{qubo} form. Furthermore, we introduce a penalty term to control the total number of H beads, $\NH$; since an unbiased minimization of $\EHP(C_t,\sv)$ has the homopolymer sequence of all H as a trivial solution. 

\begin{figure}[t]
    \centering
    \includegraphics[width=0.4\linewidth]{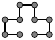}
    \includegraphics[width=0.17\linewidth]{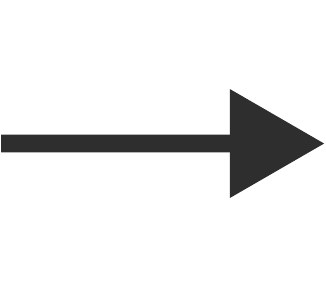}
    \includegraphics[width=0.4\linewidth]{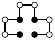}
    \caption{
    \textbf{Illustration of HP sequence optimization for $N=10$.} Given a structure (left) and an $\NH$ value, in this example 4, the task is to minimize the energy $E(\sv)$ in Eq.~(\ref{eq:E}). The solution is a sequence of H ($\medbullet$) and P ($\medcirc$) beads (right).}
    \label{fig:example_structure}
\end{figure}

As in any biophysical model based on pairwise contact interactions, the only structural information required to calculate $\EHP(C_t,\sv)$ is the contact matrix $w_{ij}$, which indicates whether two arbitrary beads $i$ and $j$ are in contact ($w_{ij}=1$) or not ($w_{ij}=0$). 
When using the HP model, a suitable choice of total energy $E(\sv)$ to minimize is given by
\begin{equation}\label{eq:E}
E(\sv)=-\sum_{1\le i<j\le N} w_{ij}s_is_j + \lambda\left(\sum_{i=1}^Ns_i-\NH\right)^2
\end{equation}
where $s_i$ describes whether bead $i$ is of type P ($s_i=0$) or H ($s_i=1$). In Eq.~(\ref{eq:E}), the
first term represents the HP interaction energy $\EHP(C_t,\sv)$, while the second term biases the total number of H-type beads toward a preset value, $\NH$. The balance between the two terms is set by the Lagrange multiplier $\lambda$. Figure~\ref{fig:example_structure} shows an example of the sequence optimization problem for $(N,\NH)=(10,4)$.
Minimizing $E(\sv)$ in Eq.~(\ref{eq:E}) can be seen as a graph bisection problem. It is a fully connected problem, but since the structure remains fixed for the sequence optimization problem its quantum formulation requires many fewer qubits than the HP folding problem~\cite{Irback:22,Linn:24} since only the bead types---not their location---need to be encoded.

The parameter $\lambda$ must be large enough for the generated sequences to acquire the desired composition, as set by $\NH$. Once above this threshold, the method's performance becomes robust to small changes in $\lambda$~\cite{Irback:24}.  
In the calculations presented below, we used $\lambda=1.1$, which worked well for all instances studied.

Below, we select one $\NH$ value for each target structure used. 
For all problem instances studied, it is possible to infer the minimum HP energy $\EHP$ by inspecting the bead-bead contacts in the target structure. 
The specific $\NH$ values used and the known minimum $\EHP$ values can be found in Appendix~\ref{app:instances}. 

\subsection{Problem-informed quantum circuits: \texorpdfstring{\glspl{qaoa}}{QAOA}} \label{sec:methods_qaoa}

In the standard \gls{qaoa}, the quantum circuit $U(\bm{\theta})$ consists of an alternating sequence of $p$ mixer and $p$ problem unitaries with the respective forms $U_M(\beta)=e^{-i\beta M}$ and $U_C(\gamma)=e^{-i\gamma C}$.  Here, $C$ encodes the cost function to be optimized, and the mixer Hamiltonian $M$ is a sum of Pauli-$X$ matrices. 
Given an initial state $\ket{\psi_0}$, an approximate solution to the optimization problem is generated by maximizing 
$\langle\psi_0| U^\dagger(\bm{\theta})CU(\bm{\theta}) |\psi_0\rangle$ over the variational parameters  
$\bm{\theta}=(\beta_1,\ldots,\beta_p,\gamma_1,\ldots,\gamma_p)$.
For large $p$ and suitably chosen parameters $\bm{\theta}$, \gls{qaoa} can be seen as a discrete version of the analog quantum annealing method~\cite{Kadowaki:98, Farhi:01}, which has been used to tackle a variety of optimization problems, including protein design~\cite{Mulligan:20, Irback:24,Panizza:24}. 

In the standard algorithm, constraints are enforced by adding soft penalty terms to the cost Hamiltonian $C$. In some cases, it is possible to restrict the quantum evolution of the system such that one or more penalty terms can be dropped. This variant, called the Quantum Alternating Operator Ansatz~\cite{Hadfield:19}, requires that the initial state, $\ket{\psi_0}$, belongs to the subspace of feasible states, and that $U_M(\beta)$ does not generate transitions from feasible to unfeasible states.   

To convert Eq.~(\ref{eq:E}) to a cost Hamiltonian $C$ that can be implemented on a quantum computer, $s_i$ is replaced by $(I_i -Z_i)/2$, where $I_i$ is the identity. The state of qubit $i$ therefore decides the letter (H or P) of amino acid $i$. As such, for a chain with $N$ amino acids, only $N$ qubits are needed.   

The problem that we wish to solve involves the constraint that the total Hamming weight of the bitstring $\bm{s}$---the number of ones in the bitstring---should equal the preset composition parameter $N_{\text{H}}$, see Eq.~(\ref{eq:E}). In this case, the penalty term becomes unnecessary if we use $XY$-mixers~{\color{magenta}\cite{Hen:16,Hadfield:19,Wang:20}}. Specifically, we consider two different $XY$-mixers, 
namely a fully connected one (called $XY$-FC below) given by $\sum_{i<j} (X_iX_j+Y_iY_j)$ and one (called $XY$-ring) where the qubits are connected in a ring, 
$\sum_{i=1}^N (X_iX_{i+1}+Y_iY_{i+1})$, where $X_{N+1}=X_1$ and $Y_{N+1}=Y_1$. For each of the two $XY$-mixers, we explore two choices of the initial state $\ket{\psi_0}$. The first is to pick some computational basis state with the desired Hamming weight $\NH$; we choose the one where the first $\NH$ bits are set to one (called BI). The second choice considered is a uniform superposition of all computational basis states with the desired Hamming weight (called DI), which is often referred to as a Dicke state~\cite{Bärtschi:19}. To prepare the Dicke states, we follow the method given in Ref.~\cite{Bärtschi:19}, for which the circuit depth scales as $\bigO(N)$ (but with $\bigO(\NH N)$ gates).  

In the following, we investigate and compare the performance of these four variants of \gls{qaoa}, all using $XY$-mixers, and standard \gls{qaoa}, with its $X$-mixer. In the latter method, the initial state $\ket{\psi_0}$ is a uniform superposition of all computational basis states (called UI). A summary
of the \gls{qaoa} variants studied can be found in Table~\ref{tab:qaoa_variants}.

\begin{table}[t]
    \caption{\label{tab:qaoa_variants} The \gls{qaoa} variants studied. We consider two $XY$-mixers with which the qubits are either fully connected ($XY$-FC) or connected in a ring ($XY$-ring). This initial state can be a uniform superposition of all computational basis states (UI), a uniform superposition of all such states with the desired Hamming weight $\NH$ (Dicke state; DI), or a single such state with Hamming weight $\NH$ ($\ket{1\ldots 10\ldots 0}$; BI). 
    }  
\begin{ruledtabular}
\begin{tabular}{ccc}
   \# & Mixer, $M$ & Initial state, $\ket{\psi_0}$ \\
\hline
I   & $X$       & UI  \\
II  & $XY$-FC   & BI  \\
III & $XY$-FC   & DI  \\
IV  & $XY$-ring & BI  \\
V   & $XY$-ring & DI  \\
\end{tabular}
\end{ruledtabular}
\end{table}

\subsection{Problem-agnostic quantum circuits: \texorpdfstring{\glspl{hea}}{HEA}} \label{sec:methods_hea}
    
With some \glspl{qaoa}, it is possible to restrict the search space to feasible solutions, but this comes at the cost of an increased circuit depth (see Sec.~\ref{sec:protein_design_QAOA}). Such a depth requirement poses a challenge for near-term quantum devices. Even standard \gls{qaoa} with its simple $X$-mixer can reach circuit depths that are difficult to optimize on current quantum hardware.

To address this issue, we also consider the \gls{hea} approach~\cite{Kandala:17}, which leads to reduced circuit depths.
In \gls{hea}, the quantum circuits typically consist of alternating layers of parametrized single-qubit rotations and entangling gates (e.g., CNOTs), 
arranged in a pattern that reflects the hardware topology. While this structure enhances compatibility with \gls{nisq} devices and allows for expressive quantum states, it lacks problem-specific encoding. 

Our \gls{hea} implementations utilize single- or two-layer \EfficientSU quantum circuits, tailored to IBM's Torino device.
Figure~\ref{fig:efficientSU2} illustrates the single-layer circuit structure for the case of $N=4$ qubits. In general, with $N$ qubits, the single- and two-layer circuits comprise, respectively, $4N$ and $6N$ variational parameters.

\subsection{Optimizing \texorpdfstring{\gls{vqa}}{VQA} parameters\label{sec:methods_paropt}}

The classical optimization of the quantum circuit parameters in \gls{vqa}s is a challenging task, known to be NP-hard~\cite{bittel_training_2021}. The presence of noise and barren plateaus~\cite{McClean:18}, where the gradient effectively vanishes, makes it necessary to find the right trade-off between size and expressivity of the circuits. To mitigate these challenges, various optimization strategies are being employed. In this paper, we consider three such strategies. 

In \gls{qaoa}, we optimize the circuits by a deterministic iterative procedure~\cite{Zhou:20}. We begin by optimizing a single-layer circuit ($p=1$), with both angles initialized to $\pi$. From the optimized parameters obtained with $p-1$ layers, an initial guess for the parameters of a $p$-layer circuit is created using interpolation~\cite{Zhou:20}.

In \gls{hea}, where the basic block (Fig.~\ref{fig:efficientSU2}) has more parameters than a single QAOA layer has, we use parameter donation between different problem instances~\cite{Langfitt:23, Lyngfelt:25, Montanez-Barrera:25}, the aim being to leverage similarities between problem instances to guide the optimizer toward promising parameter regions. 
Moving step-by-step upward in problem size, we donate optimized parameters for one circuit as initial values for the optimization of the next. Specifically, we begin with the smallest instance ($N=4$), the circuit of which is optimized with all parameters initialized randomly. These optimized parameters then serve as initial values for the parameters associated with the first four qubits in the next circuit ($N=8$). The parameters associated with the remaining four qubits are randomly initialized. This process is repeated iteratively. Note that for every instance, the optimization involves some randomly initialized parameters. Therefore, we performed 20 runs for each instance, corresponding to different realizations of the random initial values.  The optimized parameters from the best run (with the highest success probability) are transferred to the next problem instance. We do not transfer parameters between noiseless and noisy simulations to ensure that the optimization remains tailored to specific noise conditions.

Our hardware experiments with \gls{hea} are carried out using two sets of optimized parameters for a given problem instance. The sets represent warm starts, obtained as classically optimized parameters under noiseless and noisy conditions, respectively. In both cases, we take the parameters from the best performing of 20 runs.
\begin{figure}[tb]
    \centering
    \begin{quantikz}[column sep=0.2cm]
& \gate{R_y(\theta_0)} & \gate{R_z(\theta_1)} & \qw & \qw & \ctrl{1} & \gate{R_y(\theta_8)} & \gate{R_z(\theta_9)} & \meter{} \\
& \gate{R_y(\theta_2)} & \gate{R_z(\theta_3)} & \qw & \ctrl{1} & \targ{} & \gate{R_y(\theta_{10})} & \gate{R_z(\theta_{11})} & \meter{} \\
& \gate{R_y(\theta_4)} & \gate{R_z(\theta_5)} & \ctrl{1} & \targ{} & \qw  & \gate{R_y(\theta_{12})} & \gate{R_z(\theta_{13})} & \meter{} \\
& \gate{R_y(\theta_6)} & \gate{R_z(\theta_7)} & \targ{} & \qw & \qw & \gate{R_y(\theta_{14})} & \gate{R_z(\theta_{15})} & \meter{}
\end{quantikz}
    \caption{\textbf{The single-layer \EfficientSU quantum circuit used in our \gls{hea} computations, for $N=4$ qubits}. It features an entangling layer of CNOT gates in a reverse linear pattern, sandwiched between two blocks of parameterized single-qubit $R_y$ and $R_z$ rotations. Final measurements are performed in the computational basis.}
    \label{fig:efficientSU2}
\end{figure}

\subsection{Computational details\label{sec:comp_details}}
All calculations are performed in Python with NumPy~\cite{harris_array_2020} and SciPy~\cite{2020SciPy-NMeth}, and all the plots are generated with Matplotlib~\cite{hunter_matplotlib_2007}.
Quantum circuit simulations are performed using Qiskit~\cite{qiskit2024} with state vector simulators for noiseless runs and with a noise model derived from IBM's Torino backend for noisy simulations. This noise model incorporates gate errors, readout errors, and thermal relaxation effects based on the device's most recent calibration data. However, it does not account for non-Markovian effects nor crosstalk between qubits.
All quantum circuits used in this work are 2-local, meaning each term in the Hamiltonian acts on at most two qubits.
All quantum hardware executions are conducted on IBM's Torino device, which features a Heron r1 processor.
Optimization is performed using the COBYLA algorithm~\cite{cobyla_Powell1994}, with a maximum of 10000 iterations. In practice, convergence is typically achieved earlier, resulting in early termination of the optimization process. 

\subsection{Analysis\label{sec:methods_analysis}}

In Sec.~\ref{sec:results} below, we analyze the circuit depths and accuracy of the \gls{qaoa} and \gls{hea} methods when applied to the HP sequence optimization problem. We consider problem instances with unique solutions (Appendix~\ref{app:instances}), which are encoded in the ground state of the cost Hamiltonian $C$. 

For \gls{qaoa}, we limit ourselves to classical simulations of the quantum circuits. Since the parameter optimization is deterministic (Sec.~\ref{sec:methods_paropt}), we perform
one run for each problem instance. As a measure of accuracy, we take the success probability, defined as the absolute square of the overlap between the ground state of $C$ and $U(\boldsymbol{\theta})\ket{\psi_0}$.

For \gls{hea}, we perform both simulations and hardware experiments. In the simulations, using parameter donation and 20 different random number seeds (Sec.~\ref{sec:methods_paropt}), we generate 20 sets of optimized parameters for each problem instance. As a measure of accuracy, we compute the average of the success probabilities obtained with these 20 sets of optimized parameters, which is referred to as the success rate. In the hardware experiments, we determine the success rate for each of two choices of optimized parameters (Sec.~\ref{sec:methods_paropt}) by averaging over many shots. 


\section{Results}\label{sec:results}

We investigate the utility of \glspl{vqa} for identifying HP sequences that minimize the energy $\EHP$ for a given target structure and composition ($N_\text{H}$). We explore two approaches: \glspl{qaoa}, with problem-informed quantum circuits
(Sec.~\ref{sec:methods_qaoa}); and \glspl{hea}, with problem-agnostic quantum circuits (Sec.~\ref{sec:methods_hea}). 

We evaluate the \glspl{vqa} by classical simulations of the quantum circuits, both with and without noise. The former simulations use the noise model of IBM's Torino device. In the \gls{hea} case, we additionally carry out hardware experiments on the IBM Torino device.    

In Sec.~\ref{sec:protein_design_QAOA}, we present the results obtained for five \gls{qaoa} variants (Table~\ref{tab:qaoa_variants}), which use the standard $X$-mixer or one of two $XY$-mixers. With $XY$-mixers, it is possible to attain a high success probability for small instances ($N\le 16$) in noiseless simulations. However, all five \gls{qaoa} variants suffer from excessive circuit depth, which leads to a sharp decline in success probability when noise is included. 
These findings suggest that \glspl{qaoa}, in their current form, are not suitable for implementation on \gls{nisq} hardware for this problem.

In Sec.~\ref{sec:results_HEA}, we turn to the \gls{hea} approach, which is more compatible with hardware constraints due to its shallower circuit structure. In noiseless simulations, the success rates obtained with \gls{hea} are worse than those obtained with the best \gls{qaoa} variant. However, \gls{hea} has the advantage over \gls{qaoa} of being less sensitive to noise, especially when using a simple single-layer circuit structure.  
Therefore, our hardware experiments focus entirely on \gls{hea}.

\subsection{Problem-informed quantum circuits: \texorpdfstring{\glspl{qaoa}}{QAOA}}
\label{sec:protein_design_QAOA}

We evaluate the performance of five \gls{qaoa} variants  (Table~\ref{tab:qaoa_variants}), corresponding to different choices of the mixer Hamiltonian $M$ and the initial state $\ket{\psi_0}$, using as test bed the problem instances in Appendix~\ref{app:instances} with chain lengths $N\le 16$. The upper limit on problem size was needed due to rapidly growing circuit depths (see below).

The \gls{qaoa} computations reported below use quantum circuits with $p=15$ layers, unless otherwise stated. The variational parameters   ($\beta_1,\ldots,\beta_p,\gamma_1,\ldots,\gamma_p$)
were determined using an iterative procedure (Sec.~\ref{sec:methods_paropt}). 

We aim to understand how these circuit design choices affect the algorithm's effectiveness in idealized noiseless simulations and under a hardware-specific noise model. Our analysis highlights how circuit depth, mixer structure, and noise impact the success probability across different problem sizes.

\begin{figure}[t]
    \centering
    \includegraphics[width=0.99\linewidth]{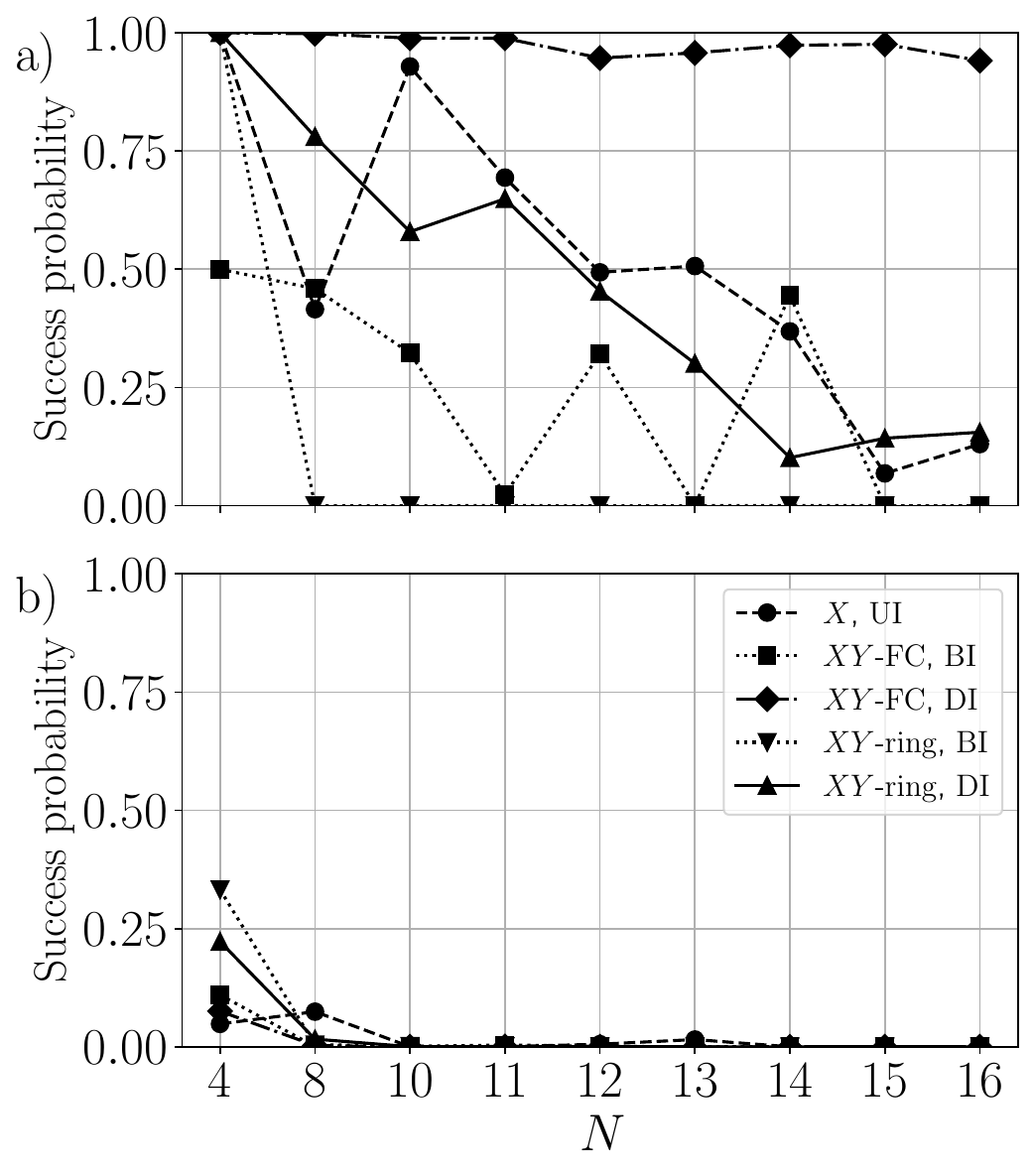}
    \caption{
    \textbf{Simulated success probability as a function of problem size ($N$) for HP sequence optimization with \glspl{qaoa}.} We consider the five \gls{qaoa} variants in Table~\ref{tab:qaoa_variants}, and the problem instances with $N\le 16$ amino acids in Appendix~\ref{app:instances}. All circuits have $p=15$ layers. The training scheme is deterministic, and the success probability is defined as the ground state probability in the final state. a) Noiseless simulations. b) Simulations using the noise model of IBM's Torino device.}
    \label{fig:qaoa_sp}
\end{figure}

Figure~\ref{fig:qaoa_sp} shows the problem size dependence of the simulated success probabilities. In the absence of noise (Fig.~\ref{fig:qaoa_sp}a), 
the best variant ($XY$-FC, DI) yields success probabilities $\geq$0.95 for all problem sizes. With the four other variants, the success probability is $<$0.5 for $N>13$. The $N$-dependence is somewhat irregular, probably in part due to the fact that we consider a single instance for each $N$. However, two trends can be seen when comparing the data for the four variants with $XY$-mixers. First, for a given choice of $XY$-mixer, the Dicke choice of initial state gives the best results. Second, for a given initial state, the fully connected $XY$-mixer gives the best results. Unfortunately, when adding noise (Fig.~\ref{fig:qaoa_sp}b), the success probability drastically drops for all five variants of \gls{qaoa} studied. 

\begin{figure}[t]
    \centering
    \includegraphics[width=0.99\linewidth]{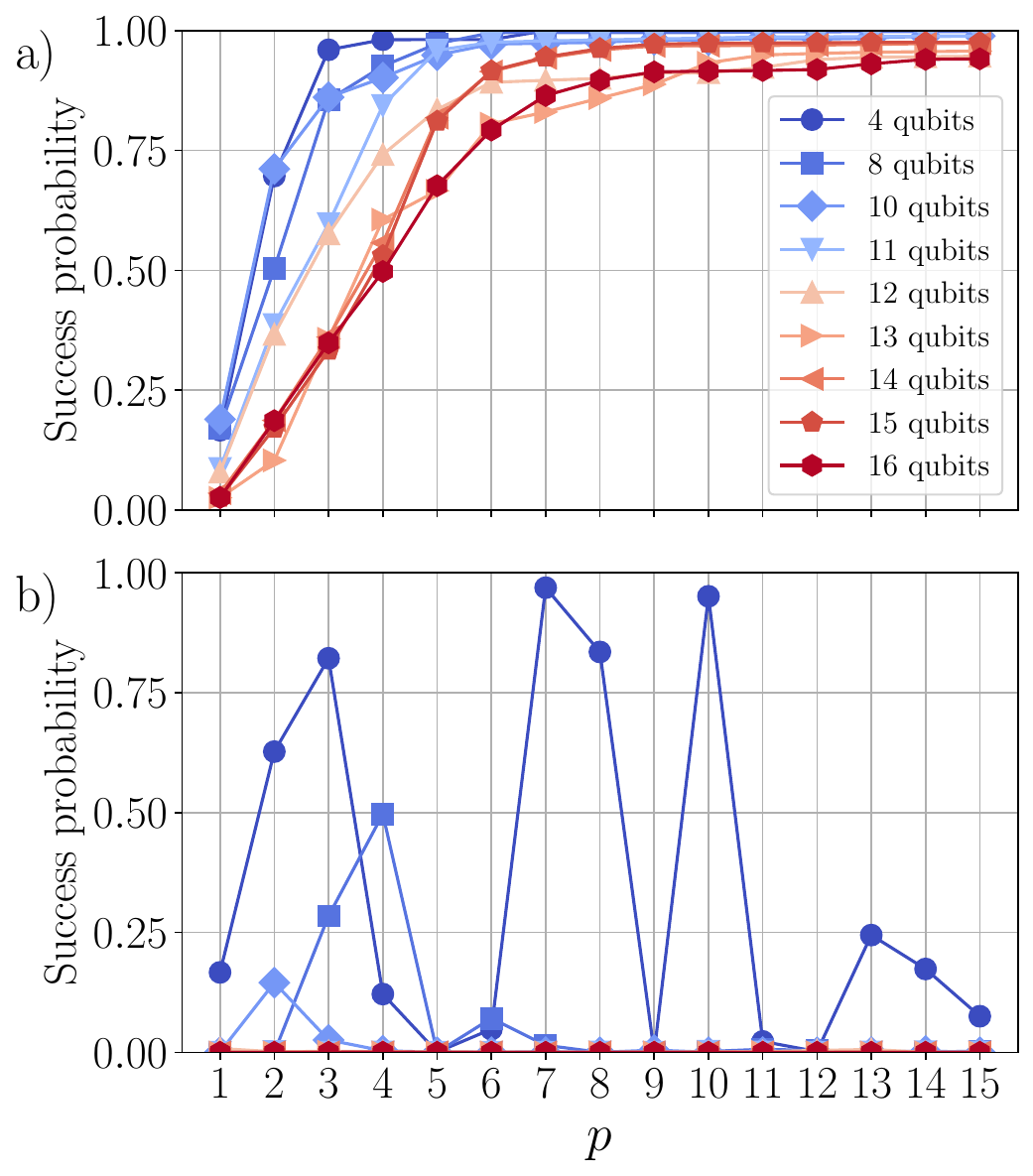}
    \caption{
    \textbf{Simulated success probability as a function of the number of QAOA layers, $p$.}
    Here, we focus on \gls{qaoa} variant III in Table~\ref{tab:qaoa_variants}, with a fully connected $XY$-mixer and a Dicke initial state. We consider the problem instances with $N\le16$ in Appendix~\ref{app:instances}. The training scheme is deterministic, and the success probability is defined as the ground state probability in the final state. a) Noiseless simulations. b) Simulations using the noise model of IBM's Torino device.
    \label{fig:qaoa_xy_fc_startinall}}
\end{figure}

To investigate whether this noise-induced drop in success probability can be avoided by reducing the number of layers, $p$, we repeated the same calculations for $1\le p<15$, focusing on the \gls{qaoa} variant 
found best in Fig.~\ref{fig:qaoa_sp}a ($XY$-FC, DI). In the absence of noise (Fig.~\ref{fig:qaoa_xy_fc_startinall}a),
the success probability increases with $p$, as expected with adequate parameter optimization. We note that in all problem instances, it is possible to reduce $p$ from 15 to $\approx$8 without any major loss in success probability. However, when adding noise (Fig.~\ref{fig:qaoa_xy_fc_startinall}b), we again observe a sharp decline in success probability, which for all problem instances with $N>10$ becomes tiny for all $p$. For the smallest instances ($N=4,8,10$), there are some values $p<15$ for which the success probability exceeds its value for $p=15$. 

\begin{figure}[tbp]
    \centering
    \includegraphics[width=0.99\linewidth]{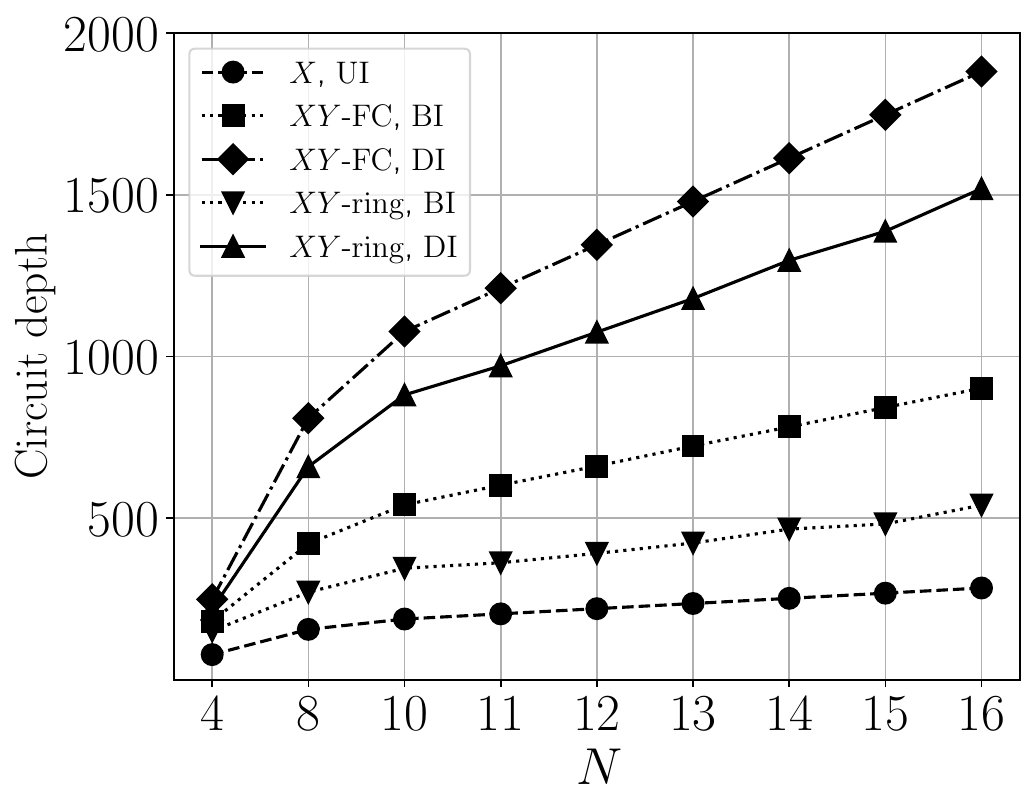}
    \caption{
    \textbf{Circuit depth as a function of problem size ($N$) for HP sequence optimization with \glspl{qaoa}.} We consider the five \gls{qaoa} variants in Table~\ref{tab:qaoa_variants}, and the problem instances with $N\le 16$ in Appendix~\ref{app:instances}. All circuits use $p=15$ layers. 
} 
    \label{fig:qaoa_depth}
\end{figure}

We attribute the poor performance of the noisy simulations to the considerable circuit depth. Figure~\ref{fig:qaoa_depth} shows how the circuit depth increases with problem size for the five \gls{qaoa} variants. The variant with a fully connected $XY$-mixer and a Dicke initial state, which performs best in terms of success probability (Fig.~\ref{fig:qaoa_sp}a), is also the one demanding most resources, with a circuit depth of $>$2000 for $N=16$. The overhead arises because both the 
mixer and the initial state are relatively complex. In contrast, the $X$-mixer yields shallower circuits but 
lower success probabilities
(Fig.~\ref{fig:qaoa_sp}a). 

Throughout the above calculations, the variational parameters were determined through an iterative procedure in $p$, where the first optimization ($p=1$) was started with both angles set to $\pi$ (Sec.~\ref{sec:methods_paropt}). To test whether the poor performance of the noisy simulations could be due to the choice of these initial values, we also explored using parameter transfer between problem instances. Here, the $p=1$ optimization was started from optimized values for a smaller instance. Despite significant overlap between low-energy regions of different $p=1$ energy landscapes (Appendix~\ref{energy_landscapes}), this parameter transfer did not yield notable improvements. This outcome suggests that the circuit depth is the dominant limiting factor.


\subsection{Problem-agnostic quantum circuits: \texorpdfstring{\gls{hea}}{HEA}}
\label{sec:results_HEA}

\begin{figure}[tbp]
    \centering    \includegraphics[width=0.95\linewidth]{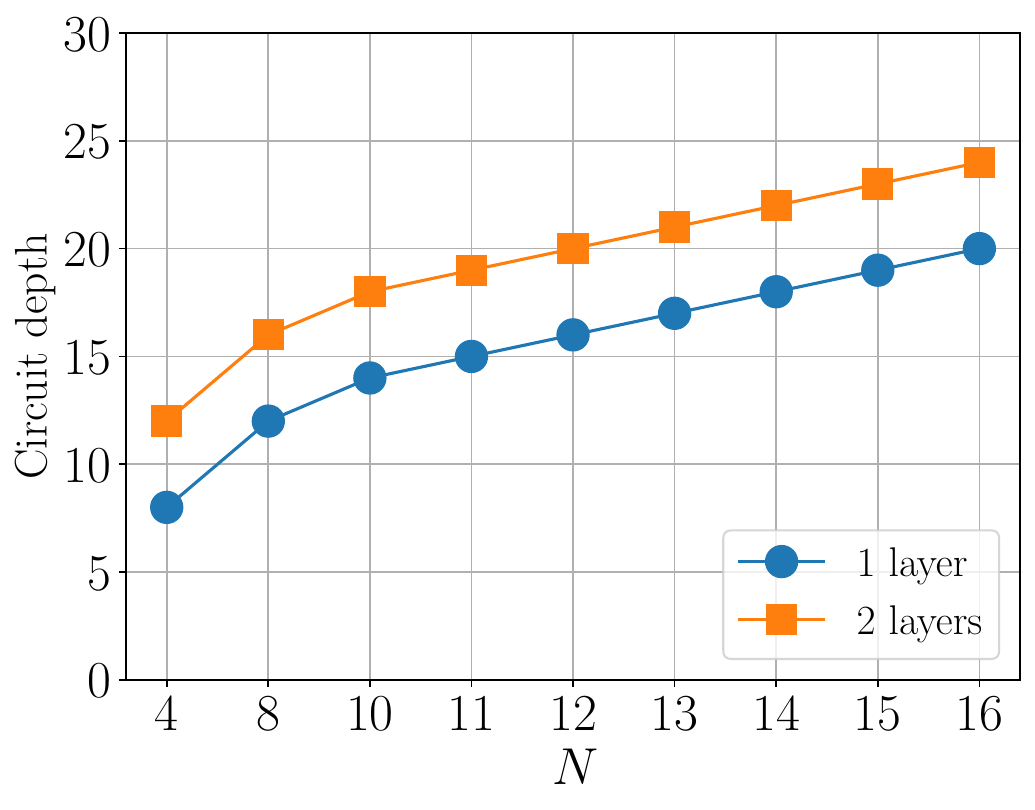}
    \caption{\textbf{Circuit depth as a function of problem size ($N$) for HP sequence optimization with \glspl{hea}.} We consider one- and two-layer \EfficientSU circuits (Sec.~\ref{sec:methods_hea}), and the problem instances with $N\le 16$ in Appendix~\ref{app:instances}.  
    }
    \label{fig:vqe_depth}
\end{figure}

To avoid the large circuit depths required by the \glspl{qaoa}, we also consider the \gls{hea} approach, with its problem-agnostic and shallower circuits. 
Specifically, we investigate the \EfficientSU ansatz, using one- and two-layer circuits (Sec.~\ref{sec:methods_hea}). Figure~\ref{fig:vqe_depth} shows how the circuit depth grows with problem size with this approach. The circuit depths are indeed significantly smaller than they are even with the most shallow \gls{qaoa} implementation using the standard $X$-mixer (Fig.~\ref{fig:qaoa_depth}). 

In the \gls{hea} computations, we use all the problem instances in Appendix~\ref{app:instances}, with chain lengths up to $N=29$. 

The \gls{hea} quantum circuits generate parameterized distributions of bitstrings. The parameters are determined to optimize the objective function, in our case, the average of the energy $E(\sv)$ in Eq.~(\ref{eq:E}). To estimate this quantity, $E(\sv)$ is computed classically for a set of bitstrings $\sv$ generated by the quantum circuit. 
We first tried optimizing the variational parameters starting from random initial values between 0 and $2\pi$, however, with poor results for large systems (data not shown). Therefore, we adopted the parameter donation scheme described in Sec.~\ref{sec:methods_paropt}.

\begin{figure*}[tbp]
    \centering
    \includegraphics[width=0.8\linewidth]{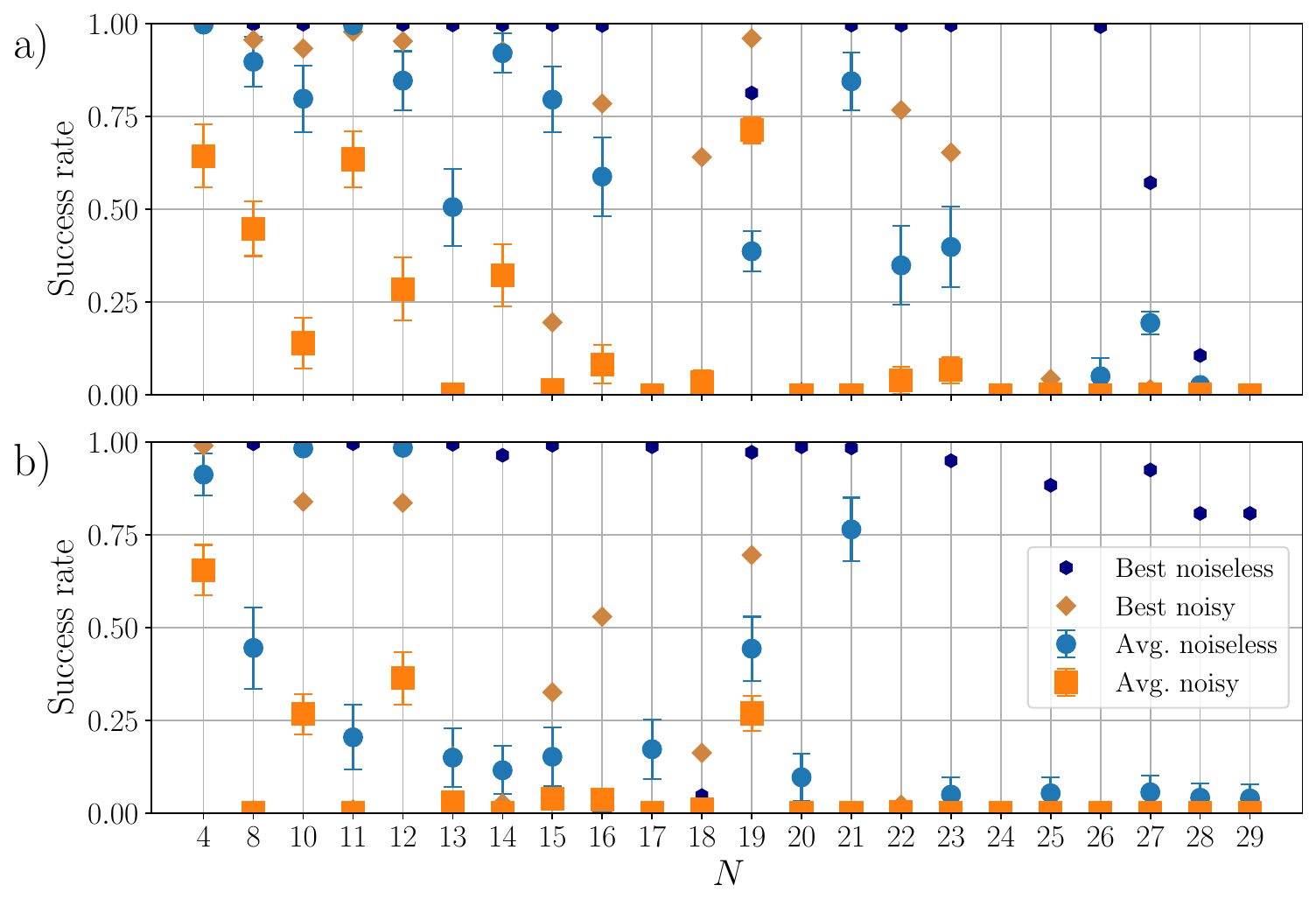}
    \caption{\textbf{Simulated success rates as a function of problem size ($N$) for HP sequence optimization with \glspl{hea}, with and without noise}. We consider the problem instances in Appendix~\ref{app:instances}. The noisy simulations used IBM's Torino noise model. For each problem instance, we generated 20 noiseless and 20 noisy runs, using different random number seeds for the parameter optimization. We show the average success probabilities over the two sets of runs as well as the highest success probabilities in the two cases. Errors bars indicate one standard error. a) One-layer \gls{hea}. b) Two-layer \gls{hea}.
    }\label{fig:vqe_success_sim}
\end{figure*}

Figure~\ref{fig:vqe_success_sim} 
shows simulated success rates  for the one- and two-layer \glspl{hea} with and without noise, as obtained using parameter donation from smaller to larger problem instances (Sec.~\ref{sec:methods_hea}). The success rate is defined as the average success probability over 20 independent runs. Additionally, Fig.~\ref{fig:vqe_success_sim} shows the success probability in the best of the 20 runs.

The results for $N\le 16$ in Fig.~\ref{fig:vqe_success_sim} can be compared with those obtained for QAOA (Fig.~\ref{fig:qaoa_sp}).  
In the noiseless case, the best \gls{qaoa} variant ($XY$-FC, DI) outperforms both one- and two-layer \gls{hea}, whose performance, however, is better than or comparable to that of the other four \gls{qaoa} variants.  

With noise, the one-layer \gls{hea} (Fig.~\ref{fig:vqe_success_sim}a) performs better than the two-layer \gls{hea} (Fig.~\ref{fig:vqe_success_sim}b), which in turn performs better than any of the \gls{qaoa} variants (Fig.~\ref{fig:qaoa_sp}b).

For the larger problem sizes ($16<N\le29$), the average success probabilities are in many cases low, especially for the two-layer circuits in the presence of noise (Fig.~\ref{fig:vqe_success_sim}b). However, the run-to-run variations are large, as indicated by significant statistical errors on the mean. 

In fact, in the noiseless case, the success probability was often close to one in at least one of the 20 runs, while vanishing in others. Therefore, the best run can have a success probability that is far above the mean (Fig.~\ref{fig:vqe_success_sim}). This effect is particularly pronounced in the two-layer case. While the average success probability is generally lower with two layers instead of one, the number of instances for which the best run has a success probability $>$0.75 is higher in the two-layer case. We believe this is because the two-layer circuit, with more parameters, is more expressive, but also more challenging to train.

The success rates in Fig.~\ref{fig:vqe_success_sim} shows a decreasing trend with increasing $N$, but the decrease is not monotonic. It is important to note monotonicity is not expected. The hardness of a problem does not depend only on $N$, but also on the choice of target structure and composition ($\NH$). The results for the $N=17$ problem illustrate this. This is the smallest $N$ for which the success rate without noise is close to zero for both one- and two-layer \gls{hea} (Fig.~\ref{fig:vqe_success_sim}). Interestingly, the same problem also proved hard for its size in a previous quantum annealing study (Fig.~4b in Ref.~\cite{Irback:24}). 

Summarizing the simulation results obtained with noise, we find that both \glspl{hea}, and especially the single-layer one, perform better than any of the \glspl{qaoa} studied. However, problem instances with $N>16$ are difficult to solve in the presence of noise, even with a single-layer \gls{hea}. 

Because of their better performance in noisy simulations, we also conducted hardware experiments for \glspl{hea}, on IBM's Torino device. These experiments were carried out using two warm-start variants, in which the variational parameters were taken from the best simulation runs with and without noise, respectively (see Fig.~\ref{fig:vqe_success_sim}).

\begin{figure*}[htbp]
\centering
    \includegraphics[width=0.8\linewidth]{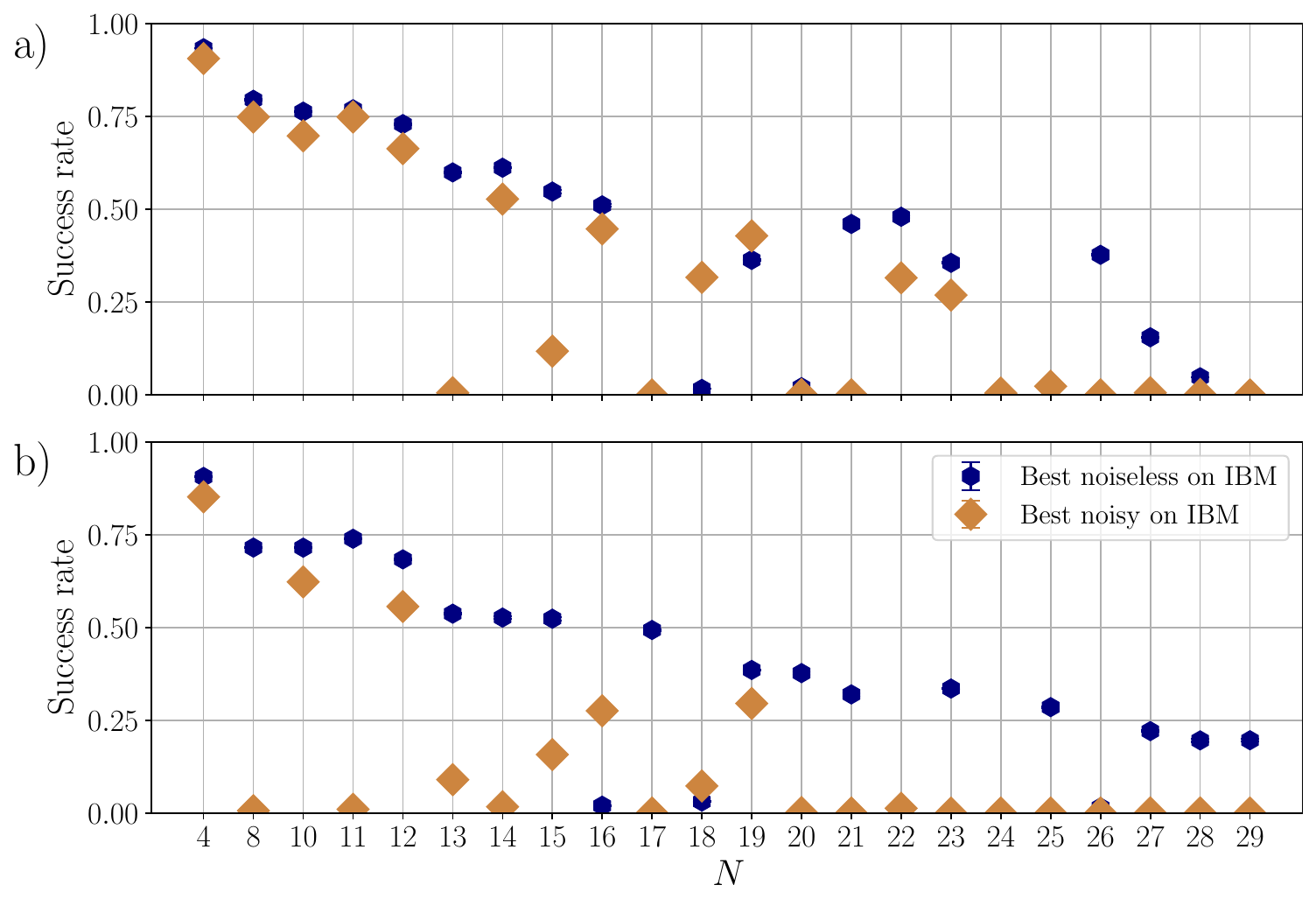}
    \caption{\textbf{Success rate of \gls{hea} on IBM's Torino device, as a function of problem size ($N$).} We use the \EfficientSU circuit ansatz (Sec.~\ref{sec:methods_hea}) and warm starts, where the parameters are taken from the best simulation runs with and without noise (Fig.~\ref{fig:vqe_success_sim}). We consider the problem instances in Appendix~\ref{app:instances}. 
    Each data point represents an average over 20 experiments, using a fixed set of optimized parameters and 10 000 shots per experiment. Statistical standard errors ($1\sigma$) are smaller than the plot symbols. a) One-layer \gls{hea}. b) Two-layer \gls{hea}.}
    \label{fig:vqe_success_qd}
\end{figure*}

Figure~\ref{fig:vqe_success_qd} shows the results of the hardware experiments for both one- and two-layer \gls{hea}. In both cases, the observed success rates tend to be higher when using parameters taken from the noiseless rather than noisy simulations. This indicates that the poorer performance of the noisy simulations (Fig.~\ref{fig:vqe_success_sim}) may in part be due to imperfectly optimized parameters.

When using two layers and parameters from the noiseless simulations, we observe success rates $>$10\% for 17 of 22 problem instances, including the three largest systems (Fig.~\ref{fig:vqe_success_qd}b). We note that the systems for which the success rate is below this threshold are precisely those for which the results of the simulations are poor (Fig.~\ref{fig:vqe_success_sim}b). The same pattern can be seen in the one-layer results as well (Figs.~\ref{fig:vqe_success_sim}a and \ref{fig:vqe_success_qd}a), although poor success rates occur somewhat more frequenly in the one-layer case (7 of 22 problem instances).
Therefore, whenever we take the parameters from a noiseless simulation that gave a high success probability, we obtain a significant success rate ($>$10\%) in the hardware experiments.

When using parameters from the best noisy simulation runs, the hardware experiments yield success rates somewhat lower than those in the simulations, possibly indicating incompleteness of the error model. Still, at a semi-quantitative level, the simulations capture the trends seen in the hardware results.  

\section{\label{sec:Discussion} Discussion}

Using \glspl{qaoa} and \glspl{hea} with, respectively, problem-informed and problem-agnostic quantum circuits, we have explored the protein sequence optimization problem, with the minimal HP model as a test bed. To this end, we first simulated classically the quantum circuits with and without noise. 

With \gls{qaoa}, it was possible to obtain high success rates under noiseless conditions, especially when using the fully connected $XY$-mixer and Dicke initial states. However, when adding noise, the performance of all five \gls{qaoa} variants studied deteriorated. We attribute this noise sensitivity to large circuit depths (Fig.~\ref{fig:qaoa_depth}). 

Note that in the circuit depth analysis for \gls{qaoa} (Fig.~\ref{fig:qaoa_depth}), we assumed a method for the preparation of Dicke states that is based on a linear nearest-neighbor topology. In practice, we have access to higher connectivity, which could allow for a more efficient preparation of Dicke states. However, since noise significantly degrades performance even when starting from the other initial states, it is unlikely that an improved Dicke state preparation would yield practical benefits under current hardware limitations.           

The shallower one- and two-layer \glspl{hea} studied, both based on the \EfficientSU ansatz, showed better noise tolerance. The most noise-tolerant among the \glspl{vqa} studied was the minimal one-layer \gls{hea}.     

Their higher noise tolerance motivated us to conduct hardware experiments with the \glspl{hea}, using the IBM Torino device. For both one- and two-layer \gls{hea}, the experimental success rates are somewhat lower than the simulated ones with the same parameters, indicating that non-negligible error sources may be missing in the error model. The latter included noise only from gate errors, gate lengths, thermal errors, and readout errors on each qubit, while neglecting temporal and correlated multi-qubit errors. Still, overall, the noisy simulations provide a semi-quantitative description of the data from the hardware experiments.    

When simulating \glspl{hea}, we found that parameter transfer between different problem instances, rather than random parameter initialization for each instance, improved the success probability for large systems. For the problem-informed \gls{qaoa} circuits, parameter transfer across instances is likely less useful. For \gls{qaoa}, we instead transferred parameters between same-instance circuits with different numbers of layers ($p$), following the iterative procedure of Ref.~\cite{Zhou:20}. In the \gls{hea} hardware experiments, we found two warm start approaches to be useful, in which the parameters were directly taken from simulations with or without noise.               
The same HP sequence optimization problem was recently addressed using analog quantum computing on a D-Wave annealer~\cite{Irback:24}. A roughly exponential decay in success rate with problem size was observed when using the purely quantum solver, with a decay rate consistent with control error estimates by D-Wave. For chain length $N=20$, success rates of $\sim$1\%-10\% were obtained~\cite{Irback:24}. As in the case of the \glspl{vqa} studied in the present paper, error mitigation seems essential in order to compete with classical optimization. The sharp deterioration in success rate with problem size is avoided with D-Wave's hybrid quantum-classical solver, whose classical part divides the problem into smaller sub-problems that are sent as queries to the QPU. With this solver, it was possible to reliably both sequence optimize and fold chains with lengths up to $N=64$~\cite{Irback:24}, competing favorably with classical Monte Carlo methods. 

Compared to D-Wave's annealers, the QAOA approach has the advantage of being flexible when it comes to the choice of mixer Hamiltonian and allows for a direct evaluation of the potential quantum advantage, as the characterization of the usage of quantum resources is straightforward. Indeed, under noiseless conditions, our best QAOA results were obtained using an $XY$-mixer rather than the standard $X$-mixer (Fig.~\ref{fig:qaoa_sp}). Unfortunately, the same QAOA variant was also the one with largest circuit depth (Fig.~\ref{fig:qaoa_depth}), which underscores the need for hardware improvements in order to unlock the potential of QAOA.

\section{\label{sec:conclusion} Conclusion and outlook}
\glspl{vqa} offer a promising approach to discrete optimization. We have implemented and tested two types of \glspl{vqa}, \glspl{qaoa} and \glspl{hea}, for sequence optimization in the HP protein model. We find that the more advanced problem-informed approach \gls{qaoa} suffers from low noise tolerance due to large circuit depths. The problem-agnostic approach \gls{hea}, with shallower circuits tailored to the hardware, has better noise tolerance. Using \gls{hea} and parameters taken from the best noiseless simulation runs, it was possible to solve the sequence optimization problem on IBM’s Torino device for 17 of the 22 problem instances studied, including the largest system with $N=29$. With access to an effective parameter determination method, it would likely be possible to obtain significant overlaps with the desired ground state for even larger problems.

Moreover, the success of D-Wave's hybrid solver with related problems suggests that gate-based methods could benefit from similar problem partitioning approaches and integration with classical heuristics. Therefore, the development and combination of current approaches with more sophisticated classical pre- and post-processing techniques could help extend the reach of gate-based algorithms to larger problem instances.

\begin{acknowledgments}
We acknowledge support from the Knut and Alice Wallenberg Foundation through the Wallenberg Center for Quantum Technology (WACQT). L.G.-{\'{A}.} further acknowledges support from the Swedish Foundation for Strategic Research (grant number FUS21-0063) and OpenSuperQ-Plus100 (101113946).

\end{acknowledgments}

\section{Code and data availability}

The code used to analyze the data is available from Ref.~\cite{GithubLinn:26}.
The data that support the findings of this study were generated by numerical simulations. The source code and parameters used to generate the simulations is publicly available~\cite{GithubLinn:26}

\appendix

\section{Problem instances}
\label{app:instances}

We solve the HP sequence optimization problem for chain lengths $4\le N\le 29$. 
Given $N$, we select one target structure and one value for the number of H beads, $\NH$. 
The choice of target structure and $\NH$ is such that (i) the problem of minimizing $\EHP$, given the target structure and $\NH$, has a unique solution, and (ii) this sequence solution has the target structure as its unique minimum $\EHP$ structure. 
Property (i) can be checked by visual inspection of the target structures, while property (ii) can be inferred from exhaustive enumerations~\cite{Irback:02,Holzgrafe:11}. 

A list of the problem instances studied can be found in Table~\ref{tab:instances}, which also shows the known minimum energies, $\EHP^{\min}$.   

\begin{table}[tb]
    \caption{\label{tab:instances} The number of amino acids, $N$, the number of H amino acids, $\NH$, and the known minimum energy, $\EHP^{\min}$, for the sequence optimization instances studied.} 
\begin{ruledtabular}
\begin{tabular}{rrr}
   $N$ &  $\NH$ & $\EHP^{\min}$ \\ 
\hline
  4  &     2 &                $-1$ \\
  8  &     4 &                $-3$ \\
  10 &     4 &                $-4$ \\
  11 &     5 &                $-4$ \\
  12 &     4 &                $-4$ \\
  13 &     8 &                $-6$ \\
  14 &     8 &                $-7$ \\
  15 &     8 &                $-7$ \\
  16 &     6 &                $-6$ \\
  17 &     6 &                $-6$ \\
  18 &     8 &                $-8$ \\
  19 &     8 &                $-8$ \\
  20 &     8 &                $-8$ \\
  21 &    10 &                $-10$ \\
  22 &    10 &                $-11$ \\
  23 &    10 &                $-10$ \\
  24 &    10 &                $-11$ \\
  25 &    13 &                $-13$ \\
  26 &    14 &                $-14$ \\
  27 &    13 &                $-13$ \\
  28 &    13 &                $-13$ \\
  29 &    15 &                $-15$ \\  
\end{tabular}
\end{ruledtabular}
\end{table}

\section{Energy landscapes for \texorpdfstring{\gls{qaoa}}{QAOA} with one layer}
\label{energy_landscapes}

\begin{figure*}[h]
  \centering
  \includegraphics[width=0.43\linewidth]{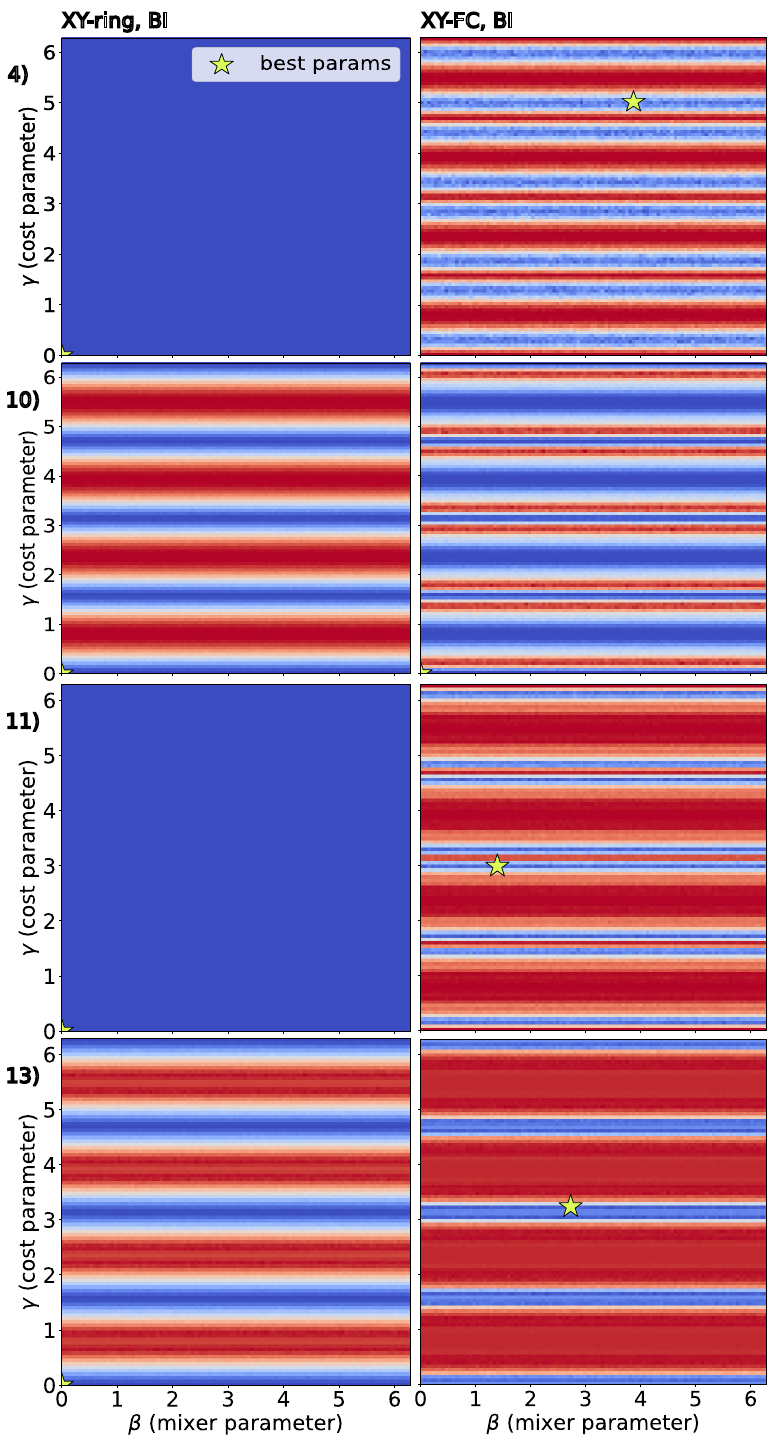}
  \caption{\textbf{QAOA energy landscapes for $p=1$ across different problem instances and circuit structures.} Each column corresponds to a different \gls{qaoa} circuit structure, while each row represents a distinct protein instance. The energy is plotted as a function of the variational parameters $\gamma$ (cost Hamiltonian) and $\beta$ (mixer Hamiltonian), with blue indicating low energy and red indicating high energy.}
  \label{fig:landscapes0}
\end{figure*}

\begin{figure*}[h]
  \centering
  \includegraphics[width=0.63\linewidth]{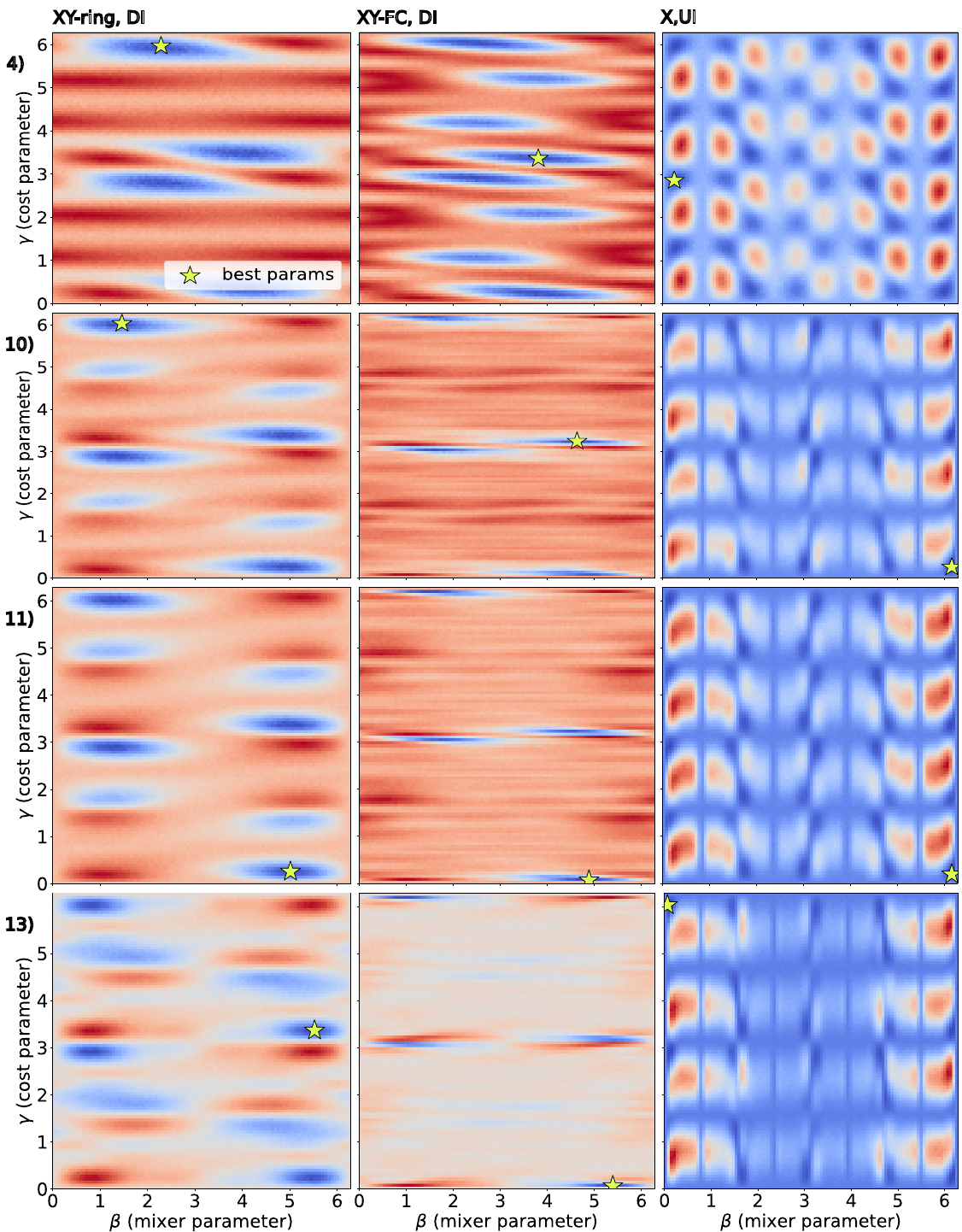}
  \caption{\textbf{QAOA energy landscapes for $p=1$ across different problem instances and circuit structures.} Each column corresponds to a different \gls{qaoa} circuit structure, while each row represents a distinct protein instance. The energy is plotted as a function of the variational parameters $\gamma$ (cost Hamiltonian) and $\beta$ (mixer Hamiltonian), with blue indicating low energy and red indicating high energy. The visual similarity between landscapes across instances suggests that optimal parameters are transferable, supporting the use of parameter donation strategies.}
  \label{fig:landscapes1}
\end{figure*}

To investigate the potential for parameter reuse across problem instances, we examine the \gls{qaoa} energy landscapes for depth $p=1$, i.e, one layer of the algorithm. Figures~\ref{fig:landscapes0} and~\ref{fig:landscapes1} show the energy as a function of the variational parameters $\gamma$ connected with the cost Hamiltonian and $\beta$ connected with the mixer Hamiltonian for two different protein instances. The similarity in landscape structure suggests that optimal parameters are not highly instance-specific, supporting the viability of parameter donation strategies. This observation motivates the use of previously optimized parameters as warm starts for related problem instances, potentially improving convergence in larger or more complex systems.

\clearpage
\bibliography{reftext}

\end{document}